\documentclass[letterpaper,preprint,preprintnumbers,amsmath,amssymb]{revtex4}

\usepackage{color}
\usepackage{graphicx}
\usepackage{bm}

\begin{document}

\includegraphics*[viewport=90 50 600 730, page=1]{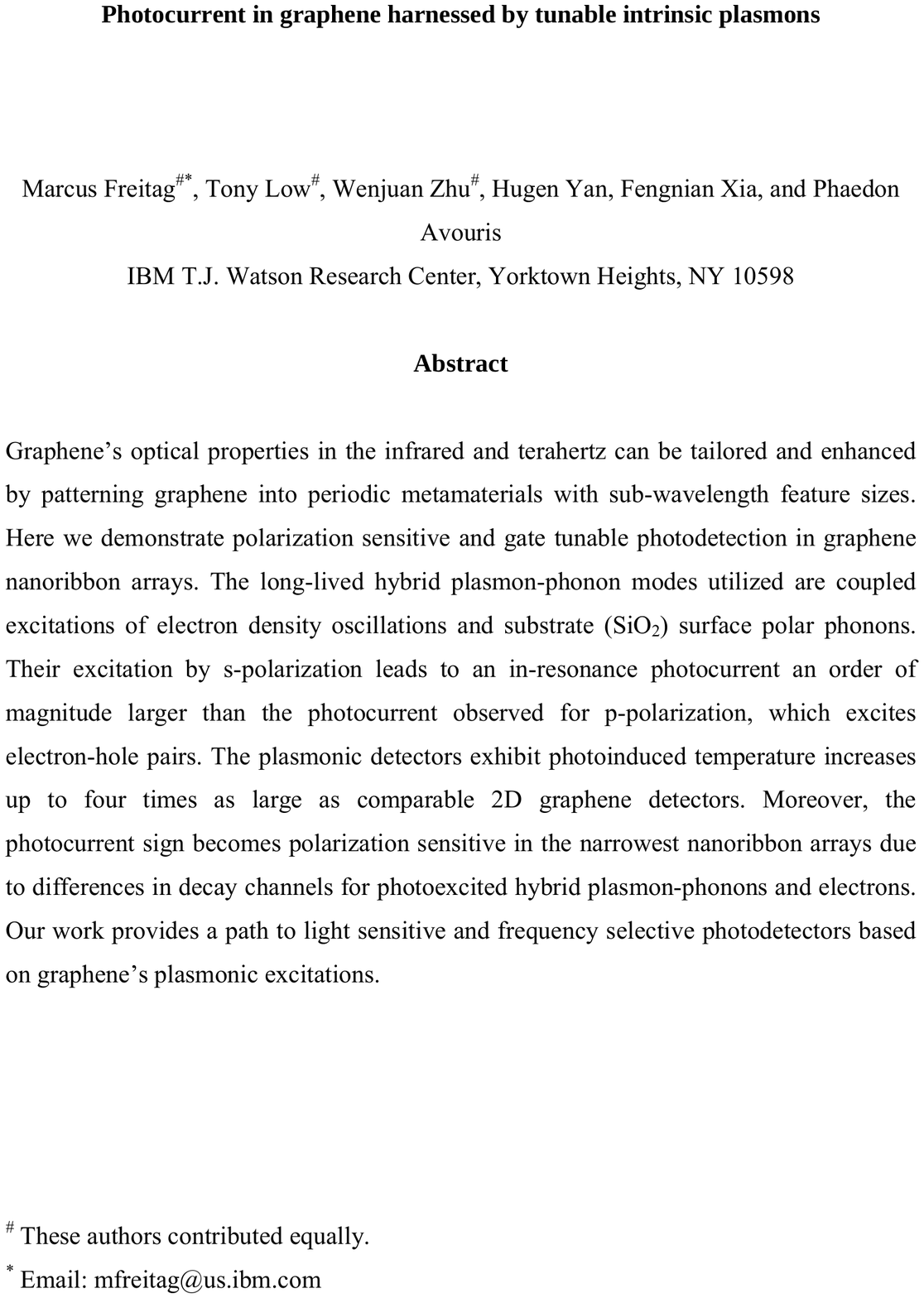}
\includegraphics*[viewport=80 50 600 730, page=2]{manuscript.pdf}
\includegraphics*[viewport=80 50 600 730, page=3]{manuscript.pdf}
\includegraphics*[viewport=80 50 600 730, page=4]{manuscript.pdf}
\includegraphics*[viewport=80 50 600 730, page=5]{manuscript.pdf}
\includegraphics*[viewport=80 50 600 730, page=6]{manuscript.pdf}
\includegraphics*[viewport=80 50 600 730, page=7]{manuscript.pdf}
\includegraphics*[viewport=80 50 600 730, page=8]{manuscript.pdf}
\includegraphics*[viewport=80 50 600 730, page=9]{manuscript.pdf}
\includegraphics*[viewport=80 50 600 730, page=10]{manuscript.pdf}
\includegraphics*[viewport=80 50 600 730, page=11]{manuscript.pdf}
\includegraphics*[viewport=80 50 600 730, page=12]{manuscript.pdf}
\includegraphics*[viewport=80 50 600 730, page=13]{manuscript.pdf}
\includegraphics*[viewport=80 50 600 730, page=14]{manuscript.pdf}
\includegraphics*[viewport=80 50 600 730, page=15]{manuscript.pdf}
\includegraphics*[viewport=80 50 600 730, page=16]{manuscript.pdf}
\includegraphics*[viewport=80 50 600 730, page=17]{manuscript.pdf}
\includegraphics*[viewport=80 50 600 730, page=18]{manuscript.pdf}
\includegraphics*[viewport=80 50 600 730, page=19]{manuscript.pdf}
\includegraphics*[viewport=80 50 600 730, page=20]{manuscript.pdf}
\includegraphics*[viewport=80 50 600 730, page=21]{manuscript.pdf}
\includegraphics*[viewport=80 50 600 730, page=22]{manuscript.pdf}
\includegraphics*[viewport=80 50 600 730, page=23]{manuscript.pdf}

\title{Supplementary Information: Photocurrent in graphene harnessed by tunable intrinsic plasmons}

\author{Marcus Freitag$^{\#}$\footnote{mfreitag@us.ibm.com}, Tony Low$^{\#}$, Wenjuan Zhu$^{\#}$,
Hugen Yan, Fengnian Xia and Phaedon Avouris}
\affiliation{IBM T.J. Watson Research Center, Yorktown Heights, NY 10598, US\\
\\(\# These authors contributed equally)}

\maketitle

\newpage

\scalebox{0.3}[0.3]{\includegraphics*[viewport=-350 -20 1400 570]{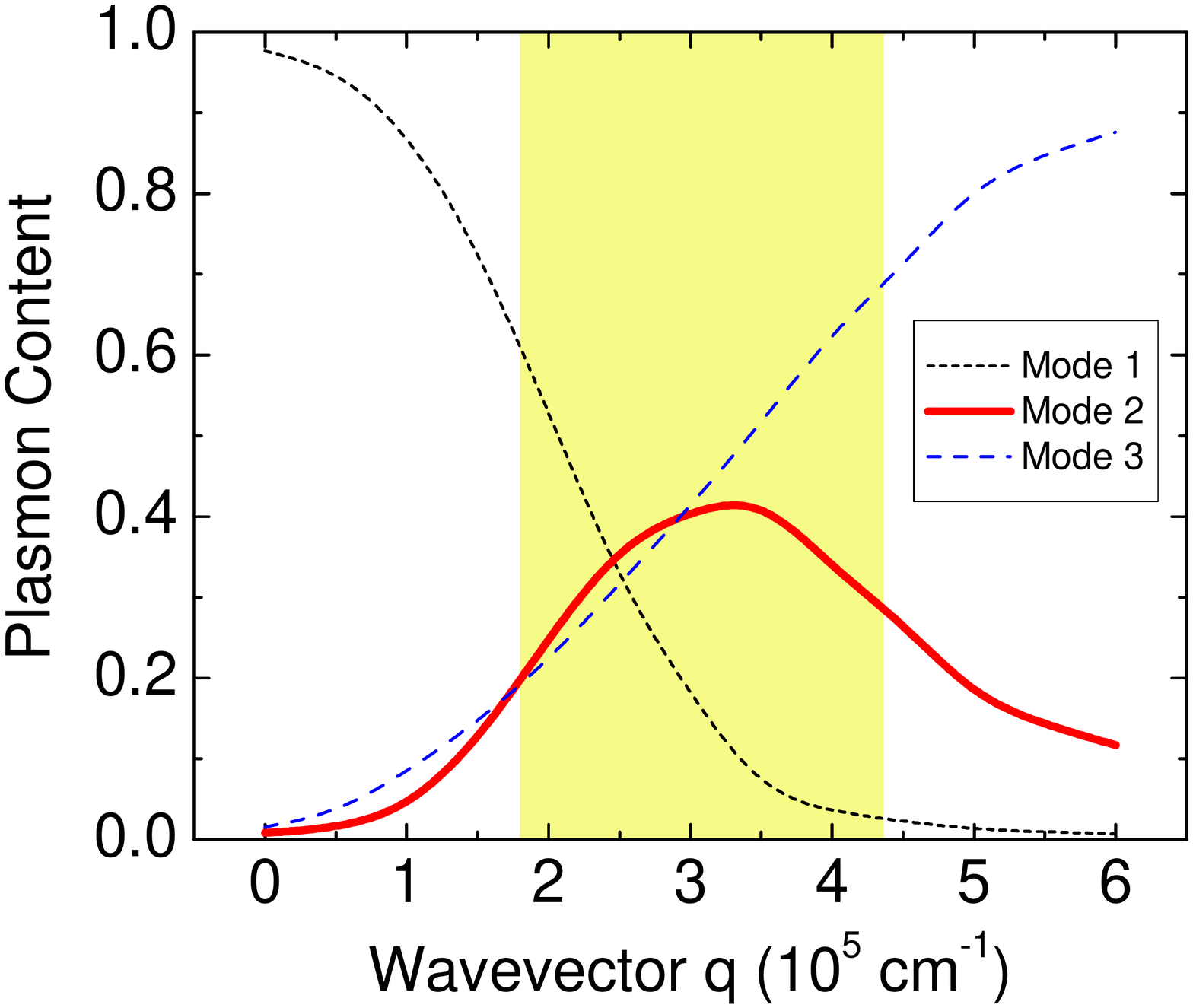}}
\textbf{Supplementary Figure S1: Plasmon content of the three hybrid plasmon-phonon modes.} Estimate based on optically measured resonance frequencies as depicted in Fig.\,3 of the main manuscript. Highlighted region indicates the range of $q$ accessed in our experiments. The plasmon-phonon modes are surface electromagnetic waves due to coupled excitations involving both collective electronic (plasmon) and ionic lattice oscillations (phonon). The resonant frequency of the coupled mode is generally different from its constituent and depends on the coupling strength. Analogous to a classical coupled harmonic oscillator problem, the nature and quality of the coupled mode (i.e. phonon- or plasmon-like) depends on its resonant frequency. For example, a coupled mode resonating at frequency close to that of the SiO$_2$ surface polar phonon exhibits narrow spectral width, inherited from the relatively long sub-picoseconds phonon lifetime.[31] Under certain physical situations, it is necessary to estimate the relative electromagnetic energy content distributed between the plasmon and phonon ``oscillators''. For example, electron scattering with coupled plasmon-phonon mode[32] requires the knowledge of plasmon and phonon content as only the latter amounts to electron's momentum relaxation.
Optoelectronic response in graphene is governed by the various energy dissipation pathways
e.g. phonons bath, contacts, substrate.[46,47 Here, we are interested in the plasmonic energy flow  into the electronic and phononic baths, which drives the optoelectronic response in graphene. Owing to the hybridization of graphene plasmon with the two surface polar phonon modes,
three plasmon-phonon coupled modes can be identified as shown in Fig.\,3 in the main text
at frequencies $\omega_1$$<$$\omega_{sp1}$ (``mode 1''), $\omega_{sp1}$$<$$\omega_2$$<$$\omega_{sp2}$ (``mode 2'')
and $\omega_3$$>$$\omega_{sp2}$ (``mode 3'').
In this work, we exploit the hybrid plasmon-phonon ``mode 2'' in our experiments.
Subsequent decay of these hybrid modes lead to dissociation into hot electron-hole pairs and substrate surface phonons,
and their relative proportion depends on the plasmon-phonon content of the particular hybrid mode.
Following Ref.\,[32], the plasmon content $\Phi_j(q)$ for mode $j$ is estimated using,
\begin{align}
\Phi_j = \frac{(\omega_j^2-\omega_{sp1}^2)(\omega_j^2-\omega_{sp2}^2)}{(\omega_j^2-\omega_{i}^2)(\omega_j^2-\omega_{k}^2)}
\tag{S1}\label{equation1}
\end{align}
where the indices $i,j,k$ are cyclical and the sum rule $\Phi_1+\Phi_2+\Phi_3=1$ holds.
Estimated $\Phi_j(q)$ for the three hybrid modes are shown above. In particular, we are interested in the plasmon content of ``mode 2'' over the range of $q$ accessed in our experiments as indicated by the highlighted region.
We see that its plasmon content exceeds $40\%$ for some intermediate $q$ which
corresponds to the $120$ and $140$\,nm ribbons.

\newpage

\scalebox{0.5}[0.5]{\includegraphics*[viewport=-100 -20 1400 450]{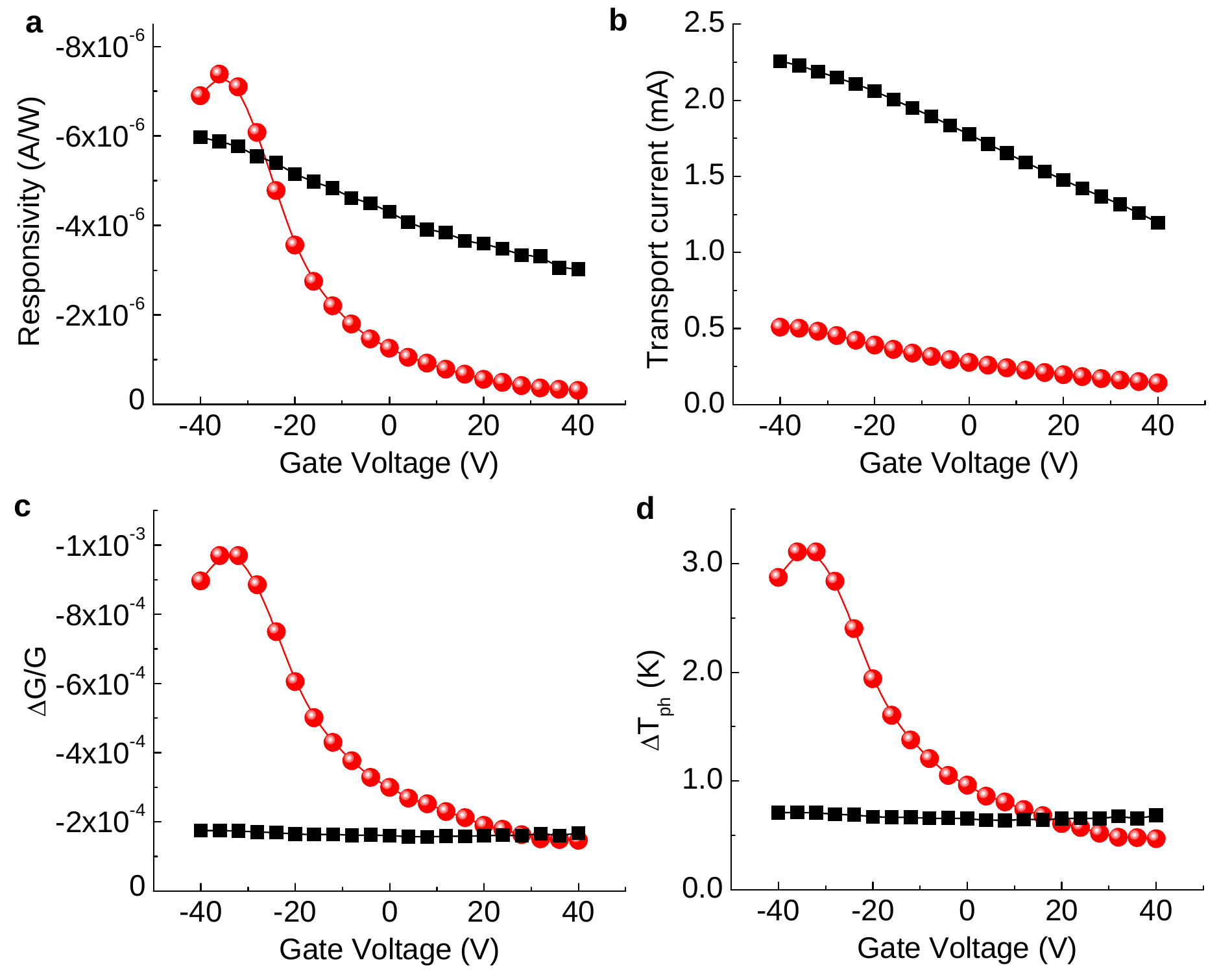}}
\textbf{Supplementary Figure S2: Comparison of GNR superlattice photodetectors with 2D graphene.} The GNR superlattice photodetector (red spheres) and 2D graphene (black squares) have identical dimensions. ($\lambda=10.6\mu$\,m, $P=66$\,mW, $V_{D}=-2$\,V). \textbf{(a)} The responsivity of $140$\,nm GNR superlattices is slightly larger than that of 2D graphene, even though it has only 1/2 fill factor. 2D graphene shows a small gradual decrease in photocurrent with gate voltage due to the decrease in transport current, while $140$\,nm GNR superlattices show a pronounced peak due to the hybrid plasmon-phonon mode. \textbf{(b)} 2D graphene has a 5 times larger transport current partly stemming from the reduced graphene coverage in the superlattice, and partly from defects or edge roughness scattering, which makes the area close to the nanoribbon edges less conductive. \textbf{(c)} By calibrating with the different transport currents and plotting $\Delta G/G$ instead, a 6 times larger response for the GNR superlattice is obtained. ($G$ is the conductance and $\Delta G=I_{ph}/V_{D}$ is the photoconductance). \textbf{(d)} The quantity $\Delta G/G$ can be translated into the lattice temperature increase $\Delta T_{ph}$ by measuring the temperature dependence of the transport current in a cryostat, from which we obtain: $\frac{\Delta G}{G \cdot \Delta T_{ph}}=-3.1 \cdot 10^{-4}$K$^{-1}$ and $-2.5 \cdot 10^{-4}$K$^{-1}$ for $140$\,nm GNRs and 2D graphene respectively. While the gate-voltage dependence of the temperature increase is flat in 2D graphene at 0.7K, it reaches $3.1$\,K in $140$\,nm GNR superlattices in resonance.

\newpage

\scalebox{0.5}[0.5]{\includegraphics*[viewport=0 -20 1200 260]{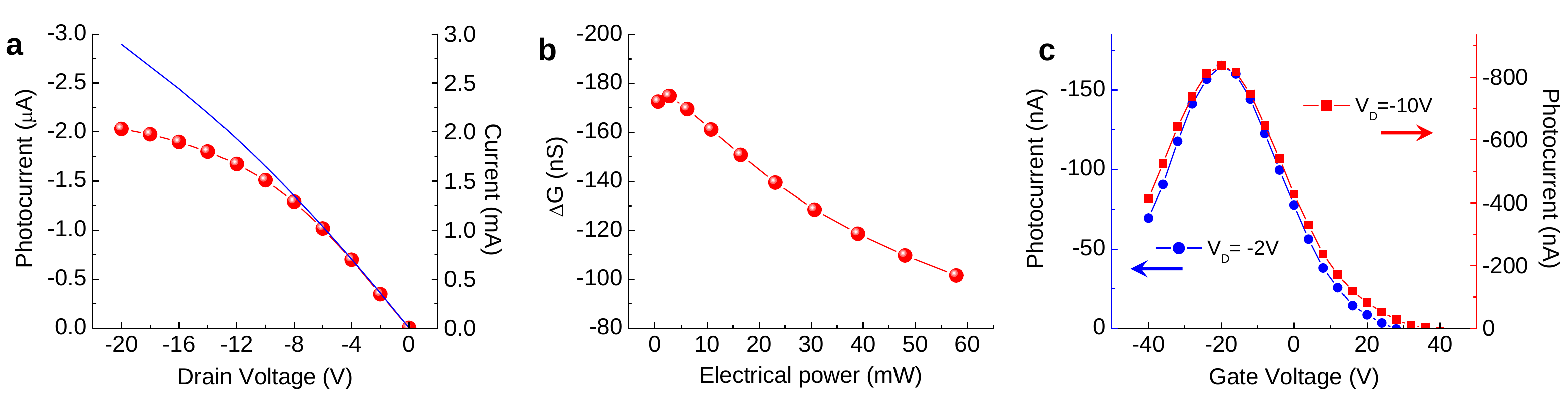}}
\textbf{Supplementary Figure S3: Effects of the drain voltage.} \textbf{(a)} We use a transport current to heat the sample (Joule heating) and measure the AC photocurrent at the same time in a 140nm GNR superlattice at $V_{G}=-40$\,V near the plasmon-phonon resonance. The temperature increase can be estimated from our previous work on the order of $\Delta T_{Joule}/P_{el} \approx$ 1K/mW for $30\mu$m devices.[42] Source-drain voltages between 2 and 20V correspond to deposited electrical powers between 1 and 60mW, and temperature increases between 1 and 60K. Photocurrent saturation sets in above -8V drain voltage, corresponding to a 10K increase in temperature. \textbf{(b)} To analyze this further, we plot the photoconductance $\Delta G=I_{ph}/V_{D}$ as a function of deposited electric power $P_{el}$. In reduction mode, $\Delta G$ is proportional to the elevated phonon temperature upon light excitation $\Delta T_{ph}$.[12] Assuming a common lattice temperature for the various participating phonon baths, the elevated phonon temperature can be described by $\Delta T_{ph}=P_{ph}/\kappa _{0}$, where $P_{ph}$ is the power absorbed by the phonon bath and $\kappa_{0}$ is the out-of-plane thermal conductance in graphene. The former is simply the absorbed power from the laser if heat dissipation via the contacts is ignored, hence relatively independent of device temperature. On the other hand, $\kappa_{0}$ increases with device temperature. Hence we expect a decreasing photo-excited phonon temperature $\Delta T_{ph}$ with increasing device temperature $\Delta T_{Joule}$ due to Joule heating. This is consistent with our experimental observation in (b). \textbf{(c)} We use a superlattice with $110$\,nm GNR width (a device we fabricated with bridges) to plot the gate voltage dependence of the photocurrent. The plasmon-phonon mode broadens by about $10\%$ or 4V in FWHM when going from -2V to -10V in drain bias. Most of this broadening will be inhomogeneous broadening due to the more pronounced potential drop along the GNR superlattice at higher drain voltage. Electron and phonon lifetimes also decrease with increasing temperature and this should lead to a broadening of the hybrid plasmon-phonon mode in combination with a reduction in peak height. The Joule heating at bias voltages of -2V and -10V translates into an increased device temperature of 1K and 15K respectively, which may lead to a small contribution to the broadening of a few percent. The associated decrease in resonance peak height can also play a contributing role in the observed decrease in photoconductance with increasing drain voltage (or device temperature) as shown in (b).

\newpage

\scalebox{0.45}[0.45]{\includegraphics*[viewport=-200 -20 1000 430]{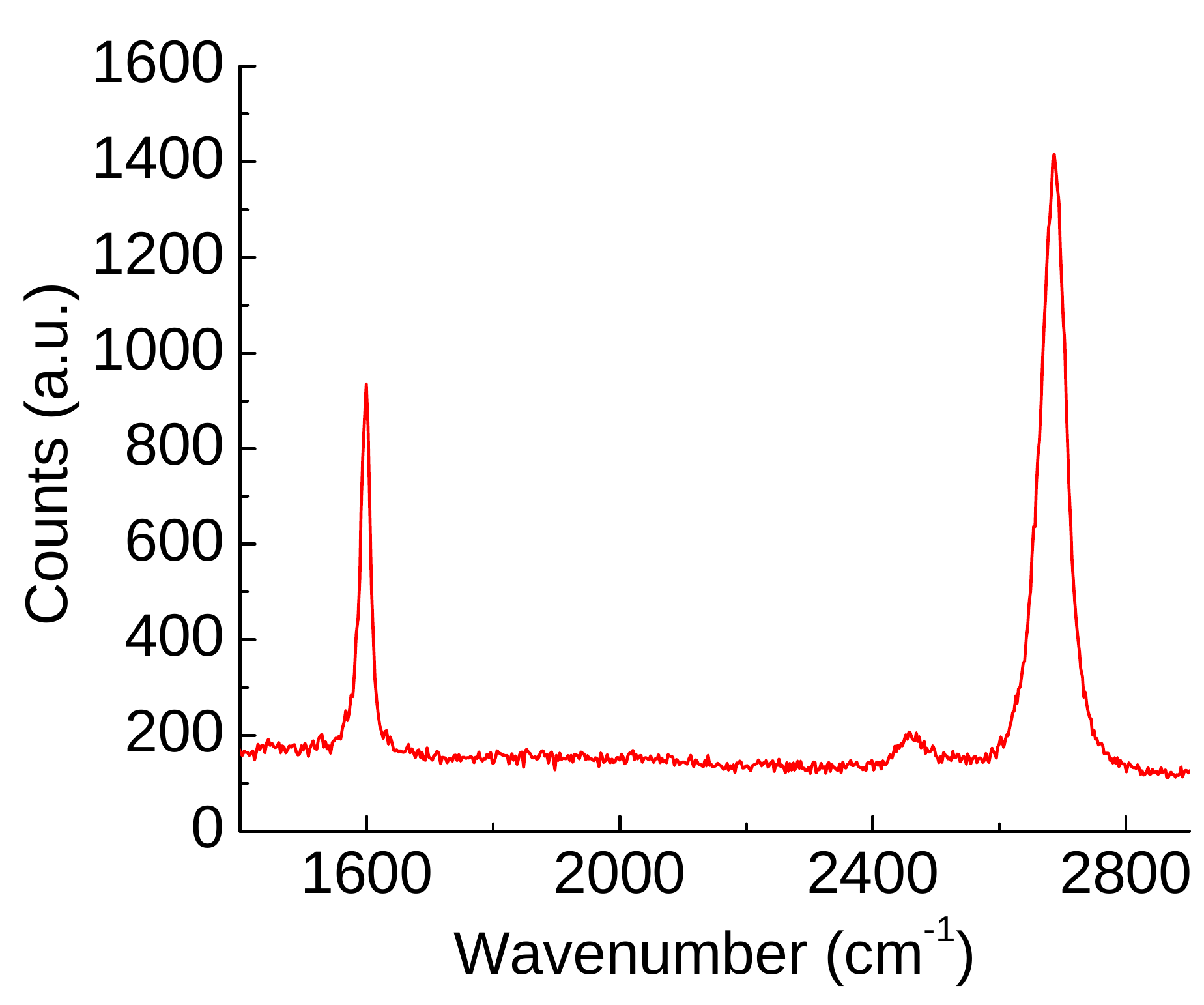}}
\textbf{Supplementary Figure S4: CVD process on copper foil produces single-layer graphene.} The Raman spectrum taken at $\lambda$$=$$532\,$nm shows a single Lorentzian G' band, indicating the presence of single-layer graphene. Our graphene is grown on copper foil, which is a process known for producing single-layer graphene almost exclusively.[45] The first layer grown passivates the copper surface and since the feedstock gas is supplied through the atmosphere, rather than being dissolved in the metal, no further layers are grown.

\newpage

\scalebox{0.42}[0.42]{\includegraphics*[viewport=0 -20 1200 300]{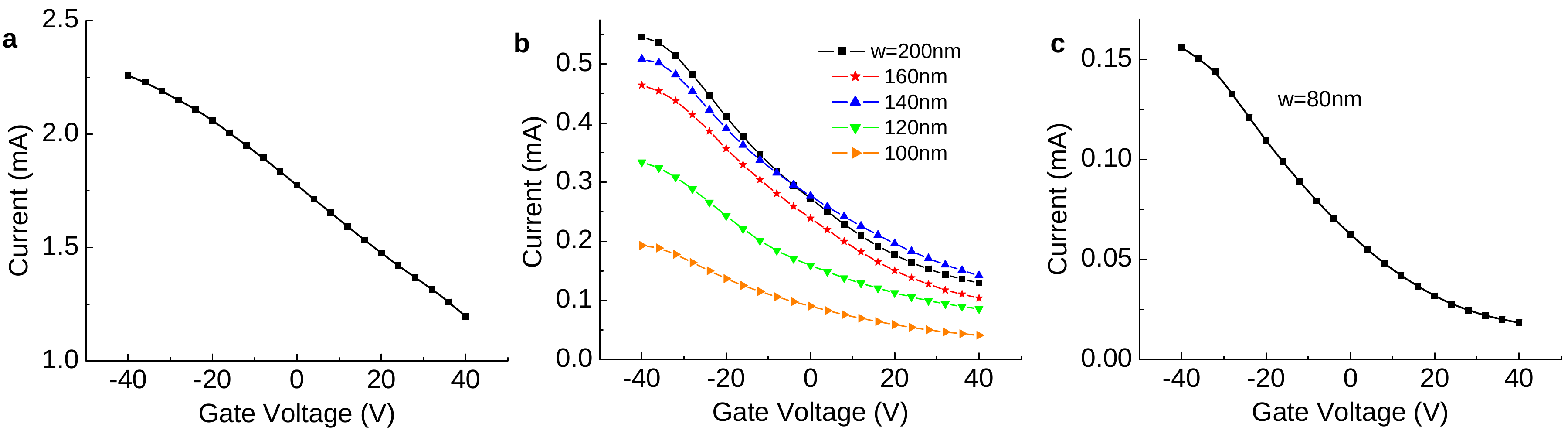}}
\textbf{Supplementary Figure S5: Electronic transport of graphene and superlattices with varying GNR width.} \textbf{(a)} Current-voltage characteristic of 2D graphene. \textbf{(b-c)} Current-voltage characteristic of GNR superlattices with GNR width as indicated. The $80$\,nm GNR superlattice contained bridges to guarantee current continuity. Electronic transport behaves p-type in air. We estimate from a combination of optical and electrical measurements a value of $52$\,V for the Dirac point gate voltage, which corresponds to a residual p-type doping of graphene of $0.33$\,eV. Superlattices with GNR width between $200$\,nm and $140$\,nm show similar transport behavior. Since we kept the fill factor (width/period) constant at $1/2$, this indicates that electronic transport is essentially 2D as in graphene. All of these devices show reduction mode photocurrent as detailed in the main text, which is consistent with the situation in 2D graphene away from the Dirac point.[12] However, when going from $140$\,nm to $120$\,nm in GNR width, the transport current is reduced by about $30\%$, and going from $120$\,nm to $100$\,nm, it is reduced by another $50\%$. (The transport current of the $80$\,nm GNR superlattice is not directly comparable). The smaller GNR superlattices have more hopping transport character, with mobility limited by edge-roughness scattering and disorder, which can be described by a transport or mobility gap,[43,48] not to be confused with a real bandgap. In these narrow GNR superlattices, the enhancement mode photoconductivity becomes more important, especially at positive gate voltages close to the Dirac point. Figure 6 of the main text shows that in superlattices with GNR width of $100$\,nm and $80$\,nm, the photocurrent becomes indeed positive in a certain gate-voltage range approaching the Dirac point, indicating that the transport gap changes the photocurrent generation mechanism. In other words, the photocarriers increase the hopping transport along the localized states in disordered GNRs, and this effect overcomes the reduction due to increased phonon scattering.

\newpage

\centerline{\textbf{Supplementary Note 1: Calculating the RPA loss function}}
\vspace{10 mm}

Here we describe the modeling of the loss function in graphene on SiO$_2$
as shown in the intensity plot of Fig.\,3 in the main manuscript.
We consider the interaction of the electronic degrees of freedom with
graphene's internal optical phonon[49,50] and that due to the surface polar phonon on SiO$_2$[40,51]. The plasmon response of graphene begins with finding the dielectric function
where a satisfactory approximation can be obtained by adding the separate contributions \emph{independently}.
An effective interaction between electrons is given by the
sum of the direct Coulomb interaction $v_c(q)=e^2/2q\epsilon_0$ where
$q$ is the wave-vector.
The two electrons interaction mediated by surface phonon and optical phonon
are denoted by
$v_{sp,\lambda}(q,\omega)$ and $v_{op}(q,\omega)$ respectively, where
their explicit expressions are given elsewhere[31].\\

The RPA expansion of the dielectric function, $\epsilon^{rpa}_T(q,\omega)$, can be expressed with this effective interaction[41,52]
\begin{align}
v_{eff}(q,\omega)=\frac{v_c(q)}{\epsilon_T^{rpa}(q,\omega)}=\frac{v_c(q)+\sum_{\lambda}v_{sp,\lambda}}{1-[v_c(q)+\sum_{\lambda}v_{sp,\lambda}]
\Pi_{\rho,\rho}^0(q,\omega)}
\tag{S2}
\end{align}
where $\Pi_{\rho,\rho}^0(q,\omega)$ is the non-interacting part (i.e. the pair bubble diagram) of
the charge-charge correlation function given by a modified Lindhard function,[22,36]
\begin{align}
\Pi_{\rho,\rho}^0(q,\omega)=-\frac{g_s}{(2\pi)^2}\sum_{nn'}\int_{\mbox{BZ}}d\bold{k}
\frac{n_F(\xi_{\bold{k}})-n_F(\xi_{\bold{k+q}})}{\xi_{\bold{k}}-\xi_{\bold{k+q}}+\hbar\omega}F_{nn'}(\bold{k},\bold{q})
\tag{S3}
\end{align}
where $n_F(\xi_{\bold{k}})$ is the Fermi-Dirac distribution function,
$F_{nn'}(\bold{k},\bold{q})$ is the band overlap function of Dirac spectrum, $g_s$ is the spin degeneracy.
While the polar surface phonons couple to the charge density operator,
the intrinsic optical phonon couple instead to the current operator. Its contribution to the dielectric function is given by $v_{op}(q,\omega)\Pi^0_{j,j}(q,\omega)$,
where $\Pi^0_{j,j}(q,\omega)$ is the current-current correlation function.
We note that from the usual charge continuity equation, $i\partial_t\hat{\rho}_{\bold{q}}=\bold{q}\cdot\hat{\bold{j}}_{\bold{q}}$,
it follows that,
\begin{align}
q^2\Pi_{j,j}(q,\omega)  = \omega^2\Pi_{\rho,\rho}(q,\omega) - v_F
\left\langle \left[\bold{q}\cdot\hat{\bold{j}}_{\bold{q}},\hat{\rho}_{-\bold{q}}\right]\right\rangle
\tag{S4}\label{chargecurr}
\end{align}
where the second term in Eq.\,\ref{chargecurr} is purely real
and $\propto q^2$ as calculated in Ref.\,[53].
The imaginary part of $\Pi_{j,j}(q,\omega)$ can be obtained just from
$\Im[\tfrac{\omega^2}{q^2}\times\Pi_{\rho,\rho}(q,\omega)]$.
Collective modes with self consistent oscillations of the carrier charge can be obtained from the zeros of the full dielectric function
\begin{align}
\epsilon_T^{rpa}(q,\omega)=\epsilon_{env}-v_{c}\Pi^0_{\rho,\rho}(q,\omega)-
\epsilon_{env} \sum_{\lambda}v_{sp,\lambda} \Pi^0_{\rho,\rho}(q,\omega)
-\epsilon_{env} v_{op}\Pi^0_{j,j}(q,\omega)
\tag{S5}\label{erpatotal}
\end{align}
where $\epsilon_{env}$ is the dielectric constant of graphene's environment.
Damping is included phenomenologically through the following modifications;
$\Pi^0_{\rho,\rho}(q,\omega)\rightarrow \Pi^0_{\rho,\rho}(q,\omega+\tau_e^{-1})$,
$v_{sp,\lambda}(q,\omega)\rightarrow v_{sp,\lambda}(q,\omega+\tau_{sp}^{-1})$
and $v_{op}(q,\omega)\rightarrow v_{op}(q,\omega+\tau_{op}^{-1})$,
where $\tau_e^{-1}$, $\tau_{sp}^{-1}$ and $\tau_{op}^{-1}$ describes
the electron, surface optical phonon and internal optical phonon lifetimes respectively.
In this work, $\tau_{sp}$ and $\tau_{op}$ are phenomenological constants
to be fitted to the experiments, while $\tau_e$ is modeled rigorously, see below.\\

Here, we discuss model description of the electron lifetime $\tau_e$.
Including relevant scattering mechanisms in our experiments, $\tau_e$ is given by,
\begin{align}
\tau_e(q,\omega) \approx \left[\tau_{0}^{-1}+\tau_{edge}(q)^{-1}+\tau_{ep}(\omega)^{-1}\right]^{-1}
\tag{S6}\label{taue}
\end{align}
where $\tau_{0}$ describes a background damping due to scattering
with impurities and $\tau_{edge}(q)\approx a/(W-W_0)^{b}$ is related to scattering off
the ribbon edges. $W$ is the ribbon's width and $W_0$ accounts for the
difference in physical and electrical widths. Experiments found this to be $\approx 28\,nm$[31].
$\tau_{0}\approx 85\,$fs as measured from the Drude response of large area, unpatterned graphene.
$a\approx 2\times 10^6$, of the order of Fermi velocity and
$b=1$ as discussed in the main text.
$\tau_{ep}(\omega)$ is electron lifetime due to
scattering with optical phonons.
It is related to the electron self-energy $\Sigma_{ep}$ via
$\tau_{ep}=\hbar/2\Im[\Sigma_{ep}]$.
According to density functional calculations, the
imaginary part of $\Sigma_{ep}$ can be approximated by,[54]
\begin{align}
\Im[\Sigma_{ep}(\omega)] = \gamma_0 \left|\hbar\omega+\hbar\omega_0+E_f\right|\times\frac{1}{2}
\left[\mbox{erf}\left(\frac{\hbar\omega-\hbar\omega_{op}}{\Delta_{ph}}\right)+\mbox{erf}\left(\frac{-\hbar\omega-\hbar\omega_{op}}{\Delta_{ph}}\right)+2\right]
\tag{S7}
\end{align}
where $\gamma_0$ describes the effective e-ph coupling and $\Delta_{ph}$
accounts for various energy broadening effects such as
the deviation from the Einstein phonon dispersion model.
They are estimated to be $\gamma_0\approx 0.018$ and
$\Delta_{ph}\approx 50\,$meV from density
function calculations.[54]\\

Using the above theory, we plot the loss function in graphene on SiO$_2$
as shown in Fig.\,3 of the main manuscript.
The calculations include interactions with the intrinsic and SiO$_2$ substrate phonons.
Graphene doping of $E_f=-0.33\,$eV and an effective $\epsilon_{env}=1.5$
is chosen to fit the plasmon modes determined from the extinction spectra.
We assume a typical $\tau_{sp}=1\,$ps while a much smaller $\tau_{op}=70\,$fs
accounts for broadening effects due to finite phonon dispersion.
Related to $v_{sp,\lambda}(q,\omega)$ and $v_{op}(q,\omega)$, we have the
frequencies of the various phonon modes at
$\omega_{op}=1580\,$cm$^{-1}$, $\omega_{sp1}=806\,$cm$^{-1}$ and $\omega_{sp2}=1168\,$cm$^{-1}$.
Their respective electron-phonon coupling parameters used are $g_0=7.7\,$eV$\AA^{-1}$,
${\cal F}^2_{sp1}=0.2\,$meV and ${\cal F}^2_{sp2}=2\,$meV.

\vspace{10 mm}

\centerline{\textbf{Supplementary References}}

\vspace{10 mm}

\noindent [46]	Low, T., Perebeinos, V., Kim, R., Freitag, M. and Avouris, P. Cooling of photoexcited carriers in graphene by internal and substrate phonons. \textit{Phys. Rev. }\textbf{B 86}, 045413 (2012).\\
\noindent [47]	Freitag, M., Low, T. and Avouris, P. Increased Responsivity of Suspended Graphene Photodetectors. \textit{Nano Lett. }\textbf{13}, 1644-1648 (2013).\\
\noindent [48]	Sols, F., Guinea, F. and Castro Neto, A. H. Coulomb Blockade in Graphene Nanoribbons. \textit{Phys. Rev. Lett. }\textbf{99}, 166803 (2007).\\
\noindent [49]	Ando, T. Anomaly of Optical Phonon in Monolayer Graphene. \textit{J. Phys. Soc. Jpn. }\textbf{75}, 124701 (2006).\\
\noindent [50]	Ishikawa, K. and Ando, T. Optical Phonon Interacting with Electrons in Carbon Nanotubes. \textit{J. Phys. Soc. Jpn. }\textbf{75}, 084713 (2006).\\
\noindent [51]	Schiefele, J., Sols, F. and Guinea, F. Temperature dependence of the conductivity of graphene on boron nitride. \textit{Phys. Rev. }\textbf{B 85}, 195420 (2012).\\
\noindent [52]	Mahan, G. D. Many particle physics.  (Springer, 2000).\\
\noindent [53]	Sabio, J., Nilsson, J. and Castro Neto, A. H. f-sum rule and unconventional spectral weight transfer in graphene. \textit{Phys. Rev. }\textbf{B 78}, 075410 (2008).\\
\noindent [54]	Park, C.-H., Giustino, F., Cohen, M. L. and Louie, S. G. Velocity Renormalization and Carrier Lifetime in Graphene from the Electron-Phonon Interaction. \textit{Phys. Rev. Lett. }\textbf{99}, 086804 (2007).

\end{document}